\documentclass[aps, prx, reprint, superscriptaddress, notitlepage, floatfix, nofootinbib]{revtex4-2}

\usepackage{amsmath, amssymb, amsfonts}
\usepackage{bm} 

\usepackage{comment}

\usepackage{graphicx}
\usepackage{booktabs}
\usepackage{caption}
\usepackage{subcaption}
\usepackage{booktabs}
\usepackage{array}
\usepackage{multirow}
\usepackage{makecell}

\usepackage{multirow}
\usepackage{array}
\newcolumntype{C}[1]{>{\centering\arraybackslash}p{#1}} 

\usepackage{float}

\usepackage{siunitx}

\usepackage[colorlinks=true, allcolors=blue]{hyperref}
\usepackage[nameinlink, noabbrev]{cleveref}
\usepackage{xcolor}

\usepackage{indentfirst}

\newcommand{\ket}[1]{\vert#1\rangle}

\newcommand{\figdir}{.}

\begin{document}

\title{Reconstructing fluid velocity fields from sparse sensors using a variational quantum algorithm}

\author{Nhat-Quang Nguyen}
\affiliation{Department of Physics, Florida State University, Tallahassee, Florida 32306, USA}
\affiliation{FSU Quantum Initiative, Florida State University, Tallahassee, Florida 32306, USA}

\author{Mohammad Mehedi Hasan Akash}
\affiliation{Mechanical and Aerospace Engineering Department, FAMU-FSU College of Engineering, Florida State University, Tallahassee, Florida 32310, USA}

\author{Kourosh Shoele}
\affiliation{Mechanical and Aerospace Engineering Department, FAMU-FSU College of Engineering, Florida State University, Tallahassee, Florida 32310, USA}

\author{Yanzhu Chen}
\email{yanzhu.chen@fsu.edu}
\affiliation{Department of Physics, Florida State University, Tallahassee, Florida 32306, USA}
\affiliation{FSU Quantum Initiative, Florida State University, Tallahassee, Florida 32306, USA}

\author{Huixuan Wu}
\email{hwu@eng.famu.fsu.edu}
\affiliation{Mechanical and Aerospace Engineering Department, FAMU-FSU College of Engineering, Florida State University, Tallahassee, Florida 32310, USA}

\date{\today}

\begin{abstract} 

Reconstructing fields governed by nonlinear partial differential equations (PDEs) from sparse measurements is a challenging task because the governing equations are strongly nonlinear and observations are available at only a few locations. Fluid velocity fields are a representative case. In this paper, we propose a variational quantum algorithm that reconstructs the solution over the entire spacetime domain at once. Rather than marching in time, the method encodes the full discrete spacetime solution in a single variational quantum state, so that all time points are optimized jointly. The cost function combines a sparse-measurement mismatch term with a physics-informed PDE violation term, letting data and the governing equation constrain the solution simultaneously. We demonstrate the method on the one-dimensional Burgers and Kuramoto--Sivashinsky equations using numerical simulations. The results suggest that variational quantum algorithms with a spacetime encoding scheme offer a compact framework for reconstructing nonlinear PDE dynamics.

\end{abstract}

\maketitle

\section{Introduction}

Reconstructing high-dimensional signals or fields from a limited number of measurements is a pervasive challenge in scientific computing, image processing, and experimental sensing~\cite{CS1donoho2006compressed, CS3candes2006near}. In many applications, the number of accessible observations is far smaller than the number of degrees of freedom required to represent the quantity of interest, which renders the inverse problem underdetermined unless additional structural assumptions are imposed. Compressed sensing addresses this regime by showing that recovery can remain accurate when the unknown admits low-complexity structure, most commonly sparsity or compressibility in a suitable representation, together with measurement processes that preserve enough information about that structure~\cite{CS1donoho2006compressed, CS2candes2006robust}. 

For fields governed by partial differential equations (PDEs), sparsity-based priors are often complemented or replaced by physics-based structure, since the governing equations impose strong constraints on admissible reconstructions. A widely used strategy is to pose sparse-sensor reconstruction as a physics-regularized inverse problem that balances agreement with measurements against consistency with the underlying dynamics. Among all, Physics-informed neural networks (PINNs) exemplify this approach by representing the unknown field with a deep neural network and optimizing a loss that couples sensor mismatch with residual penalties for the governing partial differential equation, enabling reconstruction from limited data in a range of forward and inverse fluid-mechanics problems~\cite{Raissi_2019, Cai_2021}. Physics-informed Kolmogorov--Arnold networks (PIKANs) employ a similar data-and-physics objective but replace the fixed nodal activation functions of conventional multilayer perceptrons with learnable univariate functions defined along network edges, providing an alternative representation for complex PDE solutions~\cite{Liu_KAN, PIKAN}. Neural-operator architectures such as DeepONet instead learn mappings from input functions, parameters, or sparse observations to complete solution fields. Its extension, physics-informed DeepONets, further incorporates governing-equation residuals during training, enabling reconstruction across families of PDE instances rather than optimizing a separate network for each realization~\cite{Lu_DeepONet, Wang_PIDeepONet}. As the demand for spatiotemporal resolution grows, for example in near-turbulent flows, resolving multiscale features often requires prohibitively large networks, imposing a substantial computational burden~\cite{Wang_2022, Moseley2023, Wang2024}.

Quantum computing offers an alternative route by representing and simulating solutions in an exponentially large Hilbert space. Various quantum algorithms have been proposed for solving PDEs, with the potential to outperform their classical counterparts~\cite{HHL, cerezo2021variational, Jaksch_2023, Tennie2025nonlinear, Montanaro2016, Lubasch_2020, Xu2021variational, Liu2021poisson, Kyriienko2021solving, Huang2021regression, Benedetti2021evolution, leong2022variational, Linden2022, demirdjian2022variational, Childs2021, BravoPrieto2023variationalquantum, sarma2024quantum, Wright2024NISQ, Pool_2024, Kocher_2025_HybridVQA}. To accommodate the constraints imposed by current and near-term quantum devices, several variational quantum algorithms (VQAs) have been developed that encode the discretized solution in a parameterized quantum state~\cite{Lubasch_2020, sarma2024quantum}. These approaches rely on the assumption that the solution admits a sparse representation in the Hilbert space accessible to the chosen ansatz, which can be obtained by varying circuit parameters to minimize a physics-motivated objective. Standard VQA approaches for time-dependent PDEs rely on sequential time-marching, so that the solution state is prepared or updated at successive time steps \cite{leong2022variational, Kocher_2025_HybridVQA}. More recently, spacetime-encoded VQAs have been proposed to avoid the explicit time-marching process, which is resource-intensive and error-prone~\cite{Pool_2024}. In these existing algorithms, solutions are found from a known initial condition, in which the velocity fields on the entire discrete spatial lattice are available at some starting time. However, sparse-sensor reconstruction presents a distinct challenge. The boundary condition for the governing PDE is in the form of velocity field values at some, but not all, locations throughout the time window. The lack of an initial condition precludes explicit time-marching. Solving this PDE requires recovering a globally consistent spacetime solution that simultaneously satisfies sparse measurements and nonlinear dynamics from computationally undersampled data. 

In this paper, we propose a VQA with a spacetime encoding scheme for reconstructing velocity fields from sparse, time-resolved measurements. All discrete space and time degrees of freedom are represented in a single variational quantum state, which is optimized using a cost function that combines a governing-equation residual with a sensor-data mismatch. This global representation allows measurements collected at fixed spatial locations over the full observation window to be incorporated without constructing a separate time-specific circuit for every discrete time point. Consequently, compared with conventional time-marching VQAs, the number of circuit preparations associated with the measurement residual is reduced from $\mathcal{O}(N_\mathrm{t})$ to $\mathcal{O}(1)$, where $N_\mathrm{t}$  denotes the number of time steps. In addition to this reduction in sampling overhead, the spacetime encoding scheme is compatible with the structure of the boundary condition where the available sensors are restricted to certain locations in space and provide sampled data across time. 

There are two complementary spatial representations of the velocity field. In what will be referred to as the real-space approach, the quantum-state amplitudes correspond directly to field values at discrete spatial grid points, and spatial derivatives are evaluated using finite-difference operators~\cite{Childs2021, Xu2021variational}. This construction shares the spacetime encoding and variational ansatz of Ref.~\cite{Pool_2024}, but employs a different cost function suited to reconstruction without a known initial field. In the second approach, named the Galerkin-reduced-basis approach, the field is expanded in a truncated basis adapted to the domain and boundary conditions, and the quantum state encodes the time-dependent expansion coefficients. For sufficiently smooth fields, this representation can capture the dominant spatial structure with relatively few modes and permits spatial derivatives to be evaluated analytically within the chosen basis, thereby avoiding finite-difference errors.
The Galerkin formulation is particularly attractive for smooth flow problems, whose dominant coherent structures can often be represented by relatively few modes, supporting reduced-order simulation, estimation, and control. With a basis adapted to the domain and boundary conditions, the approximation can achieve spectral convergence, faster than any algebraic order for infinitely differentiable solutions and exponential for analytic solutions, while evaluating derivatives exactly within the truncated space and providing a natural framework for stability and error analysis~\cite{Rowley2017ModelReduction,Canuto2006Spectral}.

For each representation, we construct quantum circuits to prepare the encoded field and evaluate the terms in the total cost function. For a fixed differential operator and discretization order, the number of distinct measurement circuits required for each cost-function evaluation is independent of the total number of qubits, $n=n_\mathrm{t}+n_\mathrm{s}$. Excluding state preparation by the variational ansatz and for encoding the measurement data, the additional depth of each circuit scales as $\mathcal{O}(n)$. In the real-space approach, the circuit count also depends on the order of the finite-difference stencils used to approximate the spatial derivatives. By contrast, the Galerkin-reduced-basis formulation evaluates these derivatives directly within the truncated basis, thereby eliminating finite-difference errors and the associated stencil-dependent measurement overhead.

We validate our proposed framework through numerical reconstructions of the one-dimensional Burgers and Kuramoto--Sivashinsky (KS) equations.  To demonstrate the method, we employ a problem-agnostic, hardware-efficient variational ansatz~\cite{Kandala2017, Nakaji2021, Pool_2024}. Further improvements may be obtained by incorporating information about the governing dynamics into the circuit architecture or by constructing the ansatz adaptively~\cite{Grimsley2019, Tang2021, Grimsley2023, Stadelmann2025adapt}. Empirically, we find that with a suitable ansatz depth, the reconstructed velocity field can achieve root-mean-square errors of order $10^{-2}$.  For the benchmark problems considered here, this level of reconstruction accuracy is consistently obtained using time-resolved measurements from sensors located at no more than $25\%$ of the spatial grid points, which demonstrates the recovery of globally consistent spacetime fields from spatially sparse observations.

The paper is organized as follows. In Sec.~\ref{sec:Framework}, we introduce the framework for sparse PDE field reconstruction through a spacetime-encoded VQA, followed by how to represent the solution in the real-space and Galerkin-reduced-basis approaches and how to evaluate the cost functions, respectively. We then present the quantum circuits to evaluate these cost functions on quantum devices as well as the variational ansatz to represent the discretized solution in Sec.~\ref{sec:Circuits}. Sec.~\ref{sec:numerical_simulation} contains the numerical simulation results demonstrating our sparse-measurement field-reconstruction method for velocity fields governed by the Burgers and KS equations in different viscosity regimes. Finally, we summarize our results and discuss how the algorithm can be improved in Sec.~\ref{sec:conclusion}.

\section{VQA framework for fluid velocity field reconstruction}
\label{sec:Framework}

Compressed sensing techniques enable the reconstruction of high-dimensional data from a limited number of sparse measurements. This is usually done by exploiting the sparsity property of the data, meaning that the data has a sparse representation in a certain basis (e.g., Fourier basis), and governing dynamic constraints (e.g., PDEs). With sparse representation, the number of significant entries is much smaller than the full degrees of freedom of the system, and with the PDE constraint, correlations between field values are provided. This information allows us to work with a reasonably small number of effective degrees of freedom and find a unique solution to an otherwise underdetermined system.

Using a variational ansatz $s(\vec{\theta})$ to represent the solution, where $\vec{\theta}$ are the tunable parameters, casts the reconstruction as a physics-constrained optimization problem. The optimal parameters $\vec{\theta}_*$ are obtained by minimizing a composite cost function that retains both the governing dynamics and the sparse measurements:
\begin{equation}
\label{eq:theta}
    \vec{\theta}_* = \arg \min_{\vec{\theta}} [w\, \mathcal{C}_{\mathrm{PDE}}(s(\vec{\theta}))+\mathcal{C}_{\mathrm{meas}}{(s(\vec{\theta}))]},
\end{equation}
where $\mathcal{C}_{\mathrm{PDE}}$ and $\mathcal{C}_{\mathrm{meas}}$ penalize violations of the dynamics and of the measurements, respectively. The coefficient $w>0$ is a weighting hyperparameter that balances these two constraints. Decreasing $w$ places greater emphasis on fitting the measurements, whereas increasing it gives greater weight to satisfying the governing PDE. Methods such as PINNs and VQAs demonstrate this idea with different representations of the solution~\cite{Raissi_2019, Cai_2021, Lubasch_2020, Pool_2024}.

In VQA, the variational ansatz of the solution $s(\vec{\theta})$ is encoded in a parameterized quantum circuit, with the cost function and its gradients evaluated by a quantum computer. The measured values then become the input for a classical computer to perform the optimization step and updates the ansatz parameters. This procedure continues for as many iterations as required to meet a convergence criterion. 
A key challenge is to incorporate observations from a limited set of sensor locations into a cost term that can be evaluated efficiently on quantum hardware. Encoding the complete spacetime field in a single variational quantum state allows the mismatches at all sensor locations and time steps to be evaluated jointly through $\mathcal{C}_{\mathrm{meas}}$, without preparing a separate solution state for each time step.
 
In this work, we focus on reconstructing a one-dimensional fluid velocity field $u(t,x)$ under periodic boundary conditions using sparse spatial measurements collected over a time window. The dynamics is governed by a PDE
\begin{equation}
\label{eq:nonlinearPDE}
    \frac{d}{dt} u(t,x) = \mathcal{L}\!\left[u(t,x)\right]\,u(t,x),
\end{equation}
where $x, t$ are the spatial and temporal coordinates, respectively, and $\mathcal{L}$ is a combination of linear and nonlinear operators acting on the velocity field. We can discretize Eq.~\eqref{eq:nonlinearPDE} on a spacetime grid, where we use the Euler method for the time evolution. We discuss how the time step $\Delta t$ affects the stability in Appendix~\ref{app:stable_condition} for both the Burgers and KS equations. A PDE cost function for the spacetime grid can be defined as
\begin{align}
\label{eq:Cpde}
    \mathcal{C}_{\mathrm{PDE}} &=\sum_{t}\sum_{x} [u(x,t+\Delta t) \nonumber\\
    &- \left(1+\Delta t\,\widehat{\mathcal{L}}\left[u(t,x)\right]\right)u(t,x) ]^2,
\end{align}
where $\Delta t$ is the time step, and $\widehat{\mathcal{L}}$ denotes the discrete realization of the continuous operator $\mathcal{L}$, which may involve finite-difference approximations of the spatial derivatives. Higher-order spatial and temporal discretizations can also be incorporated, although they require more complex measurement circuits.  In our study, we consider a measurement setup in which data are collected by sensors placed at a fixed set of spatial locations $\mathcal{Q}$, each recording the field values over the entire discrete time evolution. To quantify the discrepancy between the reconstructed values and these measurements, we employ the squared $\ell_2$-norm error. The corresponding cost function can be written as
\begin{equation}
\label{eq:Cmeas}
    \mathcal{C}_{\mathrm{meas}}=\sum_{t} \sum_{q\in \mathcal{Q}} \left[u(x_q,t) - u_{\mathrm{meas}}(x_q,t)\right]^2,
\end{equation}
where the sum over the index $q$ includes the set of sensor locations. Eqs.~\eqref{eq:Cpde} and \eqref{eq:Cmeas} together govern the reconstruction of the entire $u(t,x)$ field through the total cost function of $\mathcal{C}_{\rm tot}=w\mathcal{C}_{\mathrm{PDE}}+\mathcal{C}_{\mathrm{meas}}$.

To demonstrate this method, we consider two nonlinear PDEs in a one-dimensional space with the periodic boundary condition: the Burgers equation: 
\begin{equation}
\label{eq:BG_eqn}
\partial_{t}u + u\partial_{x}u = \nu \partial^{2}_{x}u,
\end{equation}
and the KS equation:
\begin{equation}
\label{eq:KS_eqn}
\partial_{t}u + u\partial_{x}u + \partial^{2}_{x}u +\nu \partial^{4}_{x}u = 0,
\end{equation}
where $\nu$ denotes the viscosity parameter in the KS and Burgers equations. The framework can be extended to higher spatial dimensions by introducing additional qubit registers to encode the corresponding spatial coordinates~\cite{Lubasch_2020, sarma2024quantum}.

\subsection{The real-space approach}
\label{subsec:real-space}
In the real-space formulation, the discretized velocity field $u(t,x)$ is amplitude-encoded in a normalized quantum state $\ket{\tilde{u}}$. The quantum register is partitioned into a time register of $n_\mathrm{t}$ qubits and a spatial register of $n_\mathrm{s}$ qubits, representing $2^{n_\mathrm{t}}$ discrete time steps and $2^{n_\mathrm{s}}$ spatial grid points, respectively. 
The quantum state can be expressed as
\begin{equation}
\label{eq:u_tilde}
    \ket{\tilde{u}} = \sum_{i=0}^{2^{n_\text{t}}-1}\sum_{j=0}^{2^{n_\text{s}}-1} \tilde{u}_{ij}\, \ket{i} \otimes \ket{j},
\end{equation}
where the binary representations of $i$ and $j$ denote the computational states of the $n_\mathrm{t}$ and $n_\mathrm{s}$ qubits, respectively.  With the spacetime grid resolution defined by $\Delta x, \Delta t$, we have 
\begin{equation}
    u(i\Delta t, j\Delta x) = \lambda_0 \tilde{u}_{ij},
\end{equation}
where $\lambda_0$ is a real normalization factor treated as a variational parameter. Since $\sum_{i,j}\lvert\tilde{u}_{ij}\rvert^2=1$, $\lambda_0$ is determined by the magnitude of the velocity field and the size of the spacetime grid. The spatial derivatives are evaluated using a first-order finite-difference approximation.

At this spatial resolution, the sensors are placed at locations $\{j\Delta x\vert j\in \mathcal{Q}\}$ for a set $\mathcal{Q}$ of indices. To evaluate $\mathcal{C}_{\mathrm{meas}}$, we encode the measurement data $u_{\mathrm{meas}}(x,t)$ in the amplitudes of another quantum state $\ket{u^{\mathrm{meas}}}$ with the same Hilbert space dimension as $\ket{\tilde{u}}$,
\begin{equation}
\label{eq:u_tilde_meas}
    \ket{\tilde{u}^{\mathrm{meas}}} = \sum_{i=0}^{2^{n_\text{t}}-1} \sum_{j=0}^{2^{n_\mathrm{s}}-1} \tilde{u}^{\mathrm{meas}}_{ij}\, \ket{i} \otimes \ket{j},
\end{equation}
where $\tilde{u}^{\mathrm{meas}}_{ij}=0$ if $j \notin \mathcal{Q}$. The amplitudes and the measurement data are related by
\begin{align}
    & u^{\mathrm{meas}}(i\Delta t, j\Delta x) = \lambda^{\mathrm{meas}} \tilde{u}_{ij}^{\mathrm{meas}}, \\
    & \lambda^{\mathrm{meas}}=\sqrt{ \sum_{i=0}^{2^{n_\text{t}}-1} \sum_{j\in{\mathcal{Q}}}|{u}^{\mathrm{meas}}_{ij}|^2}.
\end{align}
This allows us to compare the reconstructed velocity field with measurements from the state overlap using methods like the swap test. 
Preparing an arbitrary quantum state, such as Eq.~\eqref{eq:u_tilde_meas}, with specific target amplitudes can be achieved using the method outlined in \cite{Mottonen_2005_UCR_StatePrep}.

The cost function for the Burgers equation in the real-space approach is
\begin{align}
\label{eq:C_BG_phys}
    \mathcal{C}_{\mathrm{BG}}^{\mathrm{real}}= & \lambda_0^2 w \Bigg\{\sum_{i=0}^{2^{n_\mathrm{t}}-2} \sum_{j=0}^{2^{n_\mathrm{s}}-1} 
    [\tilde{u}_{i+1,j} - \tilde{u}_{i,j} \nonumber\\
    &\quad +\Delta t (\lambda_0\tilde{u}_{i,j}\nabla \tilde{u}_{i,j}-\nu \nabla^2 \tilde{u}_{i,j})
    ]^2 \Bigg\} \nonumber\\
    & +\sum_{i=0}^{2^{n_\mathrm{t}}-1}\ \sum_{j \in \mathcal{Q}}
    \left[\lambda_0 \tilde{u}_{i,j} - \lambda^{\mathrm{meas}} \tilde{u}^{\mathrm{meas}}_{i,j}\right]^2,
\end{align}
where $\nabla$ and $\nabla^2$ are discrete approximations for the spatial differential operators up to second-order errors $\mathcal{O}(\Delta{x^2})$. The operator $\nabla$ is approximated using the standard centered finite-difference formula while $\nabla^2$ is approximated using a 3-point centered stencil. The cost function for the KS equation in the real-space approach can be constructed similarly using a fourth-order spatial derivative. We include the explicit form in Appendix~\ref{app:KS_cost}.

\subsection{The Galerkin-reduced-basis approach}
\label{subsec:galerkin}
In the periodic case where $x=0$ and $x=L$ represent the same spatial point, the velocity field can be represented with a truncated Fourier series. This provides a compact alternative to the real-space encoding because, for sufficiently smooth solutions, a small number of modes can capture the dominant spatial structure. In general, the basis can be adapted to the domain geometry and boundary conditions; here, for simplicity, we use the sine-only Fourier basis, corresponding to the condition $u(t,x=0)=0$ on a periodic interval. We call this choice the Galerkin-reduced-basis approach.

When the solution is smooth, using Fourier modes is efficient and exhibits spectral convergence. In shock-dominant regimes, however, global Fourier modes can be inefficient and prone to oscillatory artifacts; in such cases, localized Galerkin variants are better suited. In particular, discontinuous Galerkin formulations and spectral-element style Galerkin bases are potential options that provide stronger robustness to sharp gradients and shocks through element-wise resolution, local limiting/filters, and more flexibility in representing non-smooth features \cite{CockburnShu2001,HesthavenWarburton2008,Tadmor1989}. In the remainder of this paper, we focus on the sine-Galerkin formulation for the smooth-flow benchmarks, while noting that these variants are natural extensions for discontinuous regimes.

When the spatial resolution uses $N+1$ grid-associated points (including $x=0$, identified with $x=L$), the number of retained sine modes is $N$. First, we assume $N=2^{n_\mathrm{s}}$ to match the dimension of the Hilbert space. 
Using $n_\text{t}$ and $n_\text{s}$ qubits to represent the spacetime distribution of the solution, we rewrite Eq.~\eqref{eq:u_tilde} to distinguish from the real-space approach,
\begin{equation}
\label{eq:b_tilde}
    \ket{\tilde{b}} = \sum_{i=0}^{2^{n_\text{t}}-1}\sum_{j=0}^{2^{n_\text{s}}-1} \tilde{b}_{ij}\, \ket{i} \otimes \ket{j},
\end{equation}
as
\begin{equation}
\label{eq:DST}
    u(i\Delta t, x) = \lambda_0 \sum_{j=0}^{2^{n_\text{s}}-1} \tilde{b}_{ij} \sqrt{\frac{2}{N+1}} \sin[(j+1)\frac{\pi}{L}x].
\end{equation}
In this representation, the spatial derivatives in $\mathcal{C}_{\mathrm{PDE}}$ are evaluated directly in the truncated basis rather than through finite differences, while $\Delta x=L/(N+1)$.  Since measurement values are defined on physical grid points, a Fourier transform is used to compare reconstructed amplitudes with sensor data, which determines the implementation of $\mathcal{C}_{\mathrm{meas}}$.

To construct a cost function to be measured on $\ket{\tilde{b}}$, we adopt the Galerkin method to project the higher modes generated by the nonlinear terms in $\mathcal{C}_{\mathrm{PDE}}$ to zero. The cost function for the Burgers equation in the Galerkin-reduced-basis approach takes the following form
\begin{align}
\label{eq:C_BG_sine}
    \mathcal{C}_{\mathrm{BG}}^{\mathrm{Galerkin}} &= \lambda_0^2 w \sum_{i=0}^{2^{n_\mathrm{t}}-2} \sum_{j=0}^{2^{n_\mathrm{s}}-1}
    \Bigg\{ \tilde{b}_{i+1,j} - \tilde{b}_{i,j} \nonumber\\
    &\quad + \Delta t \Bigg[ \nu \left(\frac{(j+1)\pi}{L}\right)^2 \tilde{b}_{i,j} \nonumber\\
    &\quad -\frac{\lambda_0}{\sqrt{2(N+1)}} \frac{(j+1)\pi}{L} \sum_{k=0}^{N-j-2} \tilde{b}_{i,k}\tilde{b}_{i,k+j+1} \nonumber\\
    &\quad +\frac{\lambda_0}{2\sqrt{2(N+1)}} \frac{(j+1)\pi}{L} \sum_{k=0}^{j-1} \tilde{b}_{i,k}\tilde{b}_{i,j-k-1} \Bigg]
    \Bigg\}^{2} \nonumber\\
    &+ \sum_{i=0}^{2^{n_\mathrm{t}}-1}\ \sum_{j \in \mathcal{Q}}
    \left[ u(i\Delta t, j\Delta x) - \lambda^{\mathrm{meas}} \tilde{u}^{\mathrm{meas}}_{i,j} \right]^2.
\end{align}
We leave the cost function for the KS equation in the Galerkin-reduced-basis approach in Appendix~\ref{app:KS_cost}.

\subsection{Variational ansatz}
\label{subsec:ansatz}

\begin{figure}[t]
    \centering
    \includegraphics[width=0.45\textwidth]
        {\figdir/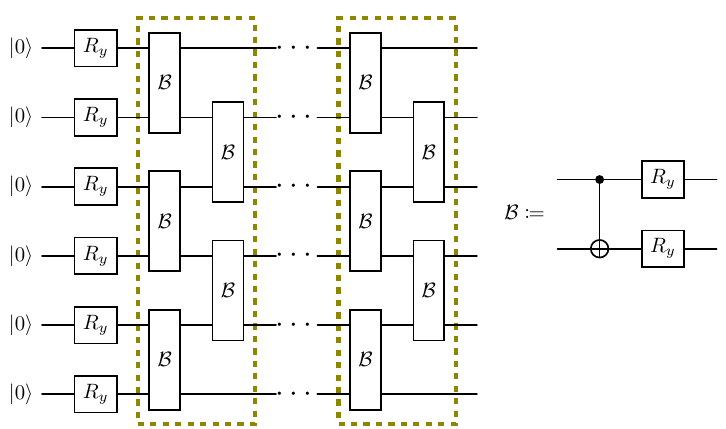}
    \caption{The hardware-efficient ansatz with repeated layers of entangling unit circuits $\mathcal{B}$ arranged in a brickwall pattern. Repeating the unit circuit $\mathcal{B}$ more times increases the expressivity of the ansatz. All the $R_y$ gates in the circuit are defined by parameters independent of each other. One dashed brown box is a brickwall layer.} 
    \label{fig:Ansatz}
\end{figure}

The variational quantum state $\ket{\tilde{u}}$ (or $\ket{\tilde{b}}$) is prepared using a variational quantum circuit on a reference state, 
\begin{equation}
    \ket{\tilde{u}(\vec{\theta})} = \mathcal{U}(\vec{\theta})\ket{\rm ref},
\end{equation}
where $\vec{\theta}$ is the set of variational parameters. The design of both the reference state $\ket{\rm ref}$ and the ansatz circuit $\mathcal{U}(\vec{\theta})$ determines the optimization landscape. 
In this work, we employ a hardware-efficient ansatz where $\ket{\rm ref}=\ket{0}^{\otimes(n_\mathrm{t}+n_\mathrm{s})}$ and $\mathcal{U}(\vec{\theta})$ consists of single-qubit rotations interleaved with several layers of CNOT gates arranged in a brickwall pattern, as shown in Fig.~\ref{fig:Ansatz}. This ansatz has been used to encode the solution to the Burgers equation on a spacetime grid when the velocity field at an initial time is available~\cite{Pool_2024}. The parameters defining the single-qubit gates are the variational parameters $\vec{\theta}$ for this ansatz. Since the target velocity field $u(x,t)$ and its modal coefficients ${\tilde b_j(t)}$ are real-valued, we restrict the ansatz to the real subspace by choosing gates so that all computational-basis amplitudes remain real throughout optimization. We enforce this by using CNOT gates and parametrized rotation gates about the $y$-axis $R_y(\theta)=e^{-i\theta Y/2}$. In our study, we define a brickwall layer as a repeating unit comprising two consecutive layers of non-overlapping CNOT gates, followed by single-qubit rotation gates. 

Hardware-efficient ansatze use gates native to most quantum devices (single-qubit rotations and native entangling gates), and yield a relatively compact and implementable circuit with moderate depth~\cite{Kandala2017, Ganzhorn2019gate, Zhuang2024}. The regular circuit structure makes training practical at small-to-moderate depth, but the problem-agnostic parameterization can still lead to a rugged optimization landscape, including barren-plateau behavior in typical VQA settings~\cite{McClean2018, Cerezo2021}. We adopt this hardware-efficient ansatz for its simplicity and for its demonstrated performance on the Burgers problem with known initial conditions~\cite{Pool_2024}. Because this ansatz is problem-agnostic and hardware-native, it offers a practical baseline with moderate depth and straightforward implementation. We expect that a problem-informed ansatz can achieve comparable accuracy with lower resource budgets, as observed in many VQA applications, including molecular ground-state calculations, where incorporating structure into the circuit can improve trainability and convergence~\cite{Grimsley2019,Tang2021, Grimsley2023, Stadelmann2025adapt}.

In Eqs.~\eqref{eq:C_BG_phys} and \eqref{eq:C_BG_sine}, the parameter $\lambda_0$ denotes the normalization factor relating the normalized register state to the field amplitude. Since the agreement with measurement data is enforced through the cost function, we treat $\lambda_0$ as a variational parameter that is determined by both the PDE constraint and the measurement-data constraint. As a reminder, a hyperparameter $w$ is used to balance the two components of the cost function and acts as a regularizer that stabilizes the optimization process. In the following numerical experiments, we select $w$ for each case based on the gradients of the two components, with values ranging from $10^{-7}$ to $10^{-4}$.

\section{Measuring the cost function}
\label{sec:Circuits}

\begin{figure}[t]
    \centering
    \includegraphics[width=\linewidth]{\figdir/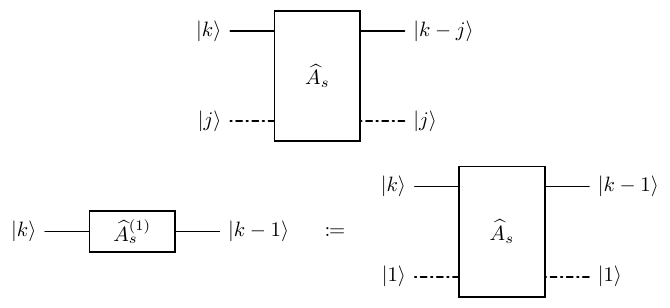}
    \caption{The circuit in the top panel illustrates the effect of the index-shifting operator $\hat{A}_s$. The state index of the top register is to be shifted and the bottom register with the dash-dotted wire encodes the value of the shift. For convenience, we define another operator $\hat{A}_s^{(1)}$, which is shifting the index by $1$, as shown in the bottom panel.}
    \label{fig:shift}
\end{figure}

To evaluate the nonlinear terms in Eq.~\eqref{eq:C_BG_phys} and Eq.~\eqref{eq:C_BG_sine}, such as $\sum_{i,j} \tilde{u}_{i+1,j}\tilde{u}_{i,j}\tilde{u}_{i,j+1}$, we use the quantum nonlinear processing unit (QNPU), developed in Refs.~\cite{Lubasch_2020, Jaksch_2023}. The general idea of QNPU is creating additional copies of the ansatz state and entangling it with the original ansatz through a set of controlled gates. The expectation value of the observable nonlinear in the state amplitudes is obtained from measuring an ancilla qubit in the Pauli $Z$ basis $\ket{0}/\ket{1}$. To evaluate terms that involve a shift of the indices, such as $\sum_{i,j}\tilde{u}_{i+1,j}\tilde{u}_{i,j}$, we take advantage of a cyclic index-shifting operator $\hat{A}_\text{s}$ whose effect is
\begin{equation}
    \hat{A}_\text{s}\ket{k}\ket{j}=\ket{k-j}\ket{j},
\end{equation}
for computational basis states $\ket{j}$ and $\ket{k}$, as shown by the defining diagram in Fig.~\ref{fig:shift}. The circuit implementation of $\hat{A}_\text{s}$ requires $\mathcal{O}(n)$ circuit depth for $n$ qubits~\cite{Cuccaro_2004}.

Because the spatial domain is periodic, sums associated with spatial-derivative terms extend over all spatial indices. In contrast, the forward-Euler discretization requires the sum over time indices to exclude the final index. We implement this restriction by adding one qubit to the time register, thereby doubling its Hilbert-space dimension, before applying the index-shifting operator~\citep{Pool_2024}. Similarly, the truncated sums over mode indices in the nonlinear terms of Eq.~\eqref{eq:C_BG_sine} are implemented by adding one additional qubit to the spatial register that encodes the mode indices.

\subsection{The real-space approach}
\label{subsec:circuit_real}

\begin{figure}[h]
    \centering
        \includegraphics[width=\linewidth]{\figdir/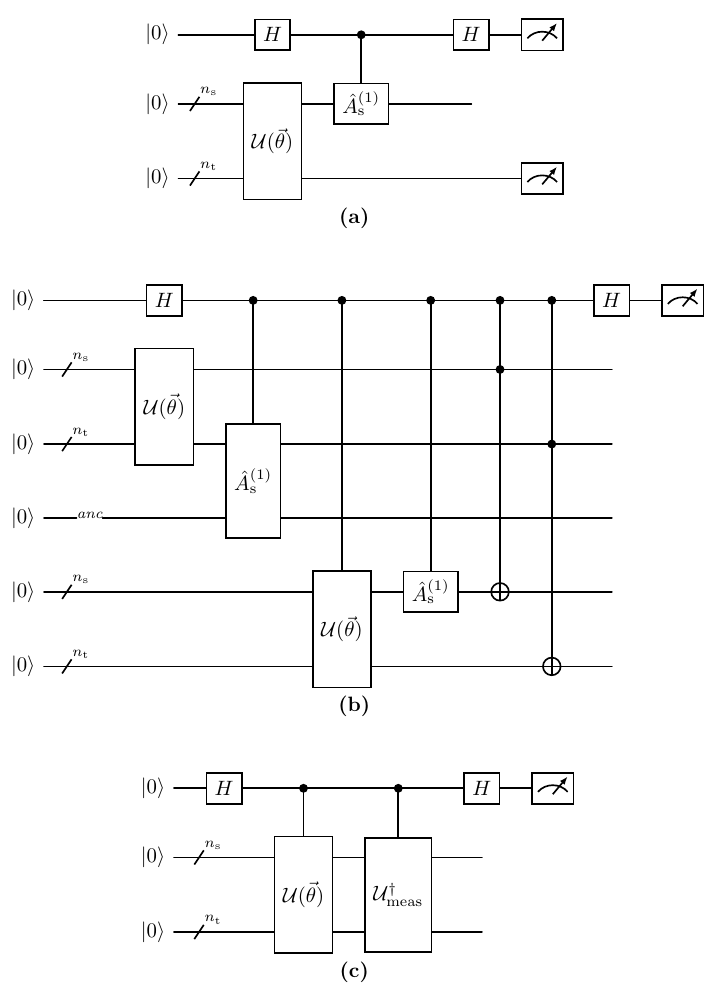}
    \caption{Circuits for evaluating representative terms in the cost function $\mathcal{C}_{\mathrm{BG}}^{\mathrm{real}}$, where $\mathcal{U}(\vec{\theta})$ denotes the variational ansatz. (a) Circuit for evaluating the linear term $\sum_{i,j} \tilde{u}_{i,j}\tilde{u}_{i,j+1}$, in which the final time index is excluded by measuring the $n_\mathrm{t}$ time registers at the end. (b) Circuit for evaluating the nonlinear term  $\sum_{i,j} \tilde{u}_{i+1,j}\tilde{u}_{i,j}\tilde{u}_{i,j+1}$, in which the final time index is excluded by introducing an additional ancilla qubit. (c) Circuit for evaluating the measurement mismatch $\sum_{i,j} \tilde{u}_{i,j}\tilde{u}^{\mathrm{meas}}_{i,j}$.}
    \label{fig:circuits_real}
\end{figure}


The PDE violation part of the cost function in Eq.~\eqref{eq:C_BG_phys} contains linear terms with two amplitudes multiplied, and nonlinear terms with three or more amplitudes multiplied. 
Not counting the ansatz circuit $\mathcal{U}(\vec{\theta})$, the depth of the circuits for evaluating the PDE violation scales linearly with the total number of qubits $n_\text{t}+n_\text{s}$. 
The number of distinct circuits required depends only on the order of the finite-difference approximation of derivatives, regardless of the number of qubits. Our implementation here takes first-order time discretization and approximates all spatial derivatives using second-order central difference, which leads to a total of 12 circuits to evaluate the PDE violation part. Higher-order approximations generate more terms to evaluate and require more circuits.

For the measurement mismatch term in Eq.~\eqref{eq:C_BG_phys}, we assume that the measurement data is encoded in a state $\ket{\tilde{u}^{\mathrm{meas}}}$. Such a state can be prepared using the method of \cite{Mottonen_2005_UCR_StatePrep}. The complexity of this auxiliary state-preparation step depends on the sparsity and the structure of the measurement data. Under the assumption that $\ket{\tilde{u}^{\mathrm{meas}}}$ has been prepared, 
the measurement mismatch term can be evaluated from the term $\sum_{i=0}^{2^{n_\mathrm{t}}-1}\ \sum_{j \in \mathcal{Q}}\vert\tilde{u}_{i,j}\vert^2$, which can be measured by one of the circuits for measuring the PDE violation part, and the state overlap between the ansatz state and $\ket{\tilde{u}^{\mathrm{meas}}}$. Since they are both real, we can use the Hadamard test to evaluate the overlap.

Fig.~\ref{fig:circuits_real} shows the circuits constructed for evaluating some representative terms in the PDE violation part of the cost function. We choose to show the following terms $\sum_{i,j} \tilde{u}_{i,j}\tilde{u}_{i,j+1}$, $\sum_{i,j} \tilde{u}_{i+1,j}\tilde{u}_{i,j}\tilde{u}_{i,j+1}$, and $\sum_{i,j} \tilde{u}_{i,j}\tilde{u}^{\mathrm{meas}}_{i,j}$, to illustrate the linear, nonlinear, and measurement mismatch terms. The total number of circuits required for evaluating Eq.~\eqref{eq:C_BG_phys} is 13.

\subsection{The Galerkin-reduced-basis approach}
\label{subsec:circuit_galerkin}

\begin{figure*}
    \centering
    \includegraphics[width=\textwidth]{\figdir/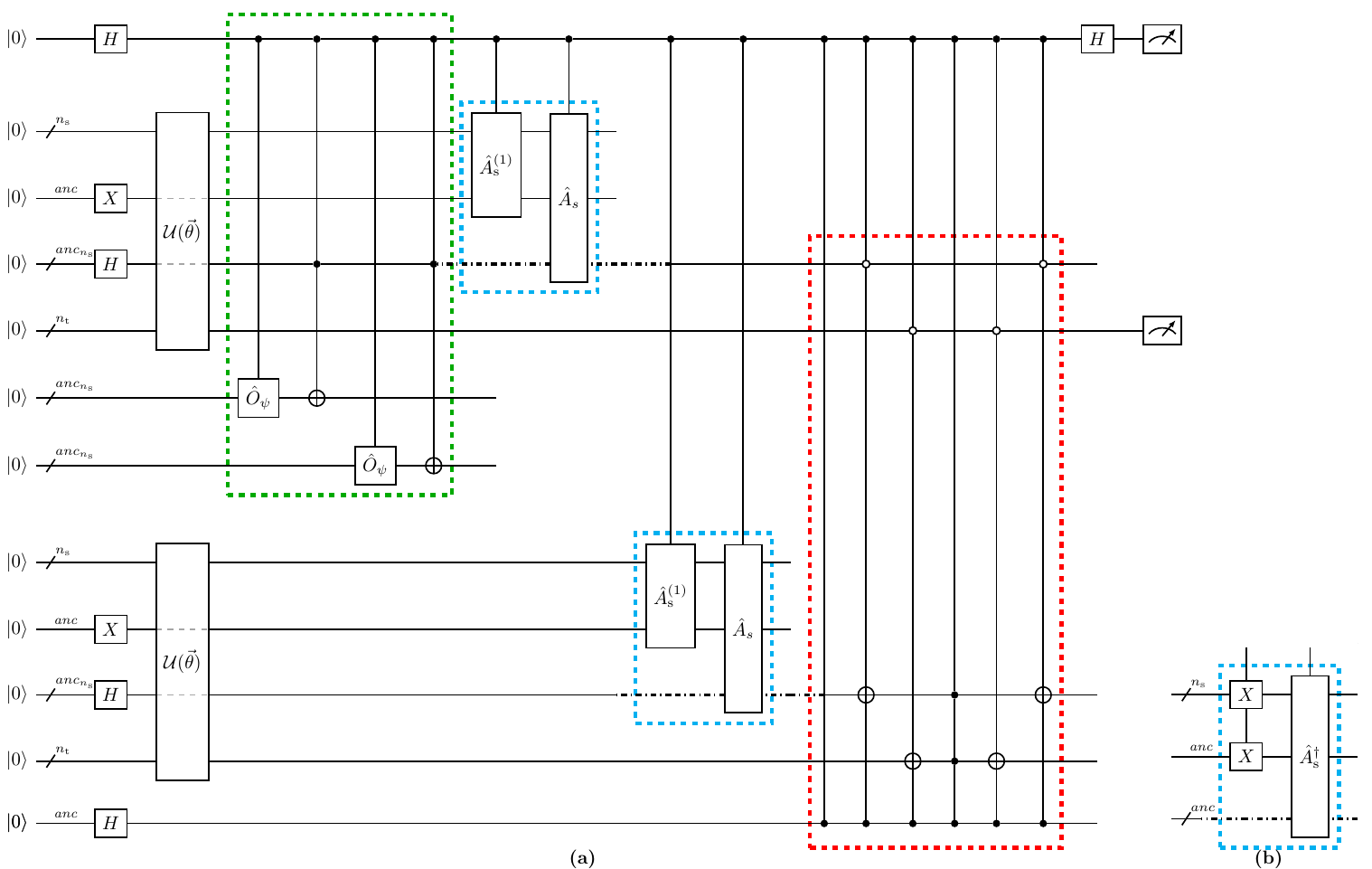}
    \caption{(a) Circuit for evaluating a representative term in Eq.~\eqref{eq:C_BG_sine}, $\sum_{i,j}\left[\sum_k(j+1)\,\tilde{b}_{i,k}\tilde{b}_{i,k+j+1}\right]^2$. Subscripts on ancilla labels indicate the number of qubits, and $\mathcal{U}(\vec{\theta})$ denotes the ansatz circuit. For each $\hat{A}_\text{s}$ box, the bottom dash-dotted wire represents the register encoding the index shift value. The green box introduces the mode-dependent factors through the auxiliary state in Eq.~\eqref{state_k}, which is prepared by the $\hat{O}_{\psi}$ operator as $\hat{O}_{\psi}\ket{0}^{\otimes{anc_{n_\text{s}}}}=\ket{\psi_\text{aux}}$. The blue box implements the index shift, and the red box performs the final step of the convolution. (b) For the circuit in (a), replacing the part in the blue box with this box enables evaluating $\sum_{i,j}\left[\sum_k(j+1)\,\tilde{b}_{i,k}\tilde{b}_{i,j-k-1}\right]^2$ instead.}
    \label{fig:convolution}
\end{figure*}

In the Galerkin-reduced-basis approach, Eq.~\eqref{eq:C_BG_sine} differs from Eq.~\eqref{eq:C_BG_phys} in the measurement mismatch term and the terms resulting from spatial derivatives in the PDE violation part. Below we discuss how to evaluate such different terms and show the circuit diagrams for some representative terms.

The nonlinear terms in the PDE lead to convolution and the spatial derivatives translate into mode-dependent multiplicative factors in Eq.~\eqref{eq:C_BG_sine}. As an example, we show how to use QNPU to evaluate a term, $\sum_{i,j}\left[\sum_k(j+1)\,\tilde{b}_{i,k}\tilde{b}_{i,k+j+1}\right]^2$, in Fig.~\ref{fig:convolution}. To evaluate the convolution, we introduce an $n_\mathrm{s}$-qubit ancilla prepared in the state $\ket{+}^{\otimes n_\mathrm{s}}$ to encode the shift in the space index. Under the Galerkin method, higher modes generated from nonlinear terms are projected out, leading to a truncation of the shifted indices to prevent aliasing. We introduce an ancilla qubit in the space register to realize this truncation. The index shift is implemented using the blue box in Fig.~\ref{fig:convolution}(a).

The mode-dependent factors can be handled using QNPU with auxiliary quantum states on an $n_\mathrm{s}$-qubit ancilla encoding the mode-dependent factors, such as 
\begin{equation}
\label{state_k}
\ket{\psi_\text{aux}}=\frac{1}{\mathcal{K}} \sum_{j=0}^{2^{n_\mathrm{s}}-1} (j+1)\ket{j},    
\end{equation}
where $\mathcal{K}=\sqrt{\sum_{j=0}^{ 2^{n_\mathrm{s}} -1}(j+1)^2}$ is the normalization factor. 
Such states, whose amplitudes are real and polynomial functions of the computational basis index, can be prepared with circuit depth $\mathcal{O}(\log n_\mathrm{s})$ and $\mathcal{O}(n_\mathrm{s})$ ancilla qubits using the method developed in Ref.~\cite{logarithmic_depth}. When higher powers of the computational basis index $j$ are required in the factors, such as $(j+1)^2$ and $(j+1)^4$, there are two ways to introduce such factors. If additional qubits are available, a QNPU with an extra $n_\mathrm{s}$-qubit ancilla prepared in the state $\ket{\psi_\text{aux}}$ can achieve this. Alternatively, we can prepare an auxiliary state encoding the amplitudes directly using the method in Ref.~\cite{logarithmic_depth}. 
In the convolution, the mode-dependent multiplicative factor depends on the shift in the index. This dependence is realized by a controlled gate in the green box in Fig.~\ref{fig:convolution}(a). The final squared convolution term is obtained by an extraction circuit (the red box in Fig.~\ref{fig:convolution}(a)). Note that the controlled gates can be implemented with depth linear in the number of qubits by using ancilla qubits (not shown in the diagram). 
With these constructions, the PDE violation part can be evaluated using 14 distinct circuits, whose depth scales linearly with the total number of qubits $n_\text{t}+n_\text{s}$. 

\begin{figure}[htbp]
    \centering
        \includegraphics[width=\columnwidth]{\figdir/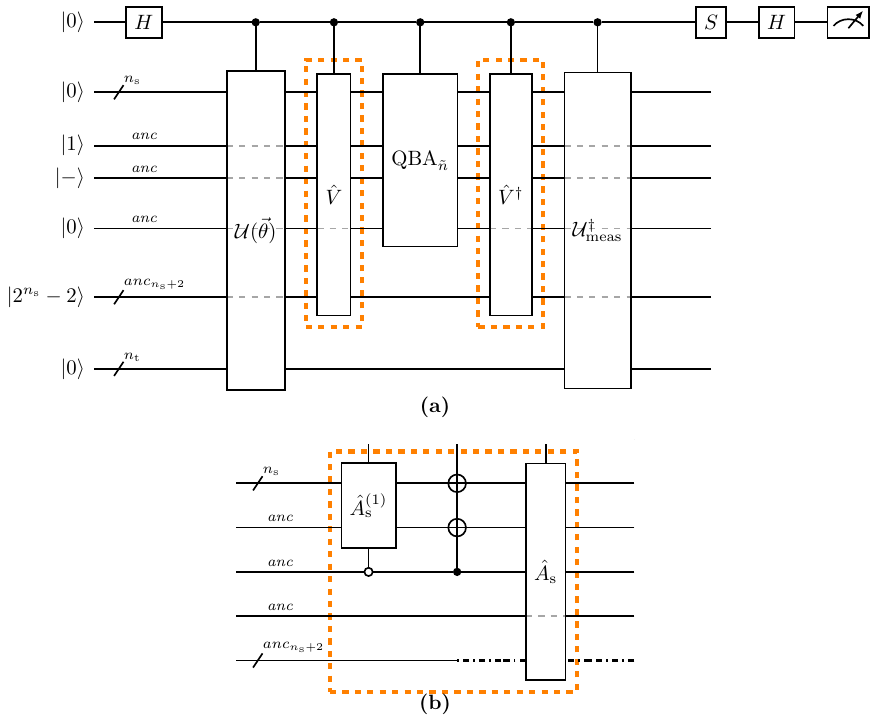}
    \caption{(a) Circuit for evaluating the measurement mismatch term in Eq.~\eqref{eq:C_BG_sine}, where $\tilde{n}=2^{n_\mathrm{s}+1}+2$ is the QBA transform length. The ancilla qubits prepared in $\ket{1}$ and $\ket{-}$ are used to enlarge the Hilbert space, preprocessing the state for QBA. Another ancilla is used for QBA. (b) Decomposition of the $\hat{V}$ circuit block. The bottom dash-dotted wire represents the register encoding the shift index used in the cyclic shift operator $\hat{A}_s$.}
    \label{QBA_measurement}
\end{figure}

To evaluate the measurement mismatch term, a discrete sine transform must be applied to map the amplitudes of the ansatz state to rescaled velocities in the real space. To implement this transform on a quantum device, we use the fact that a discrete sine transform in a domain of length $L$ is equivalent to a Fourier transform in the odd extension of the domain to length $2L$. This means that the Hilbert space encoding the spatial modes must be doubled in order to apply the transform. In addition, because the register state in Eq.~\eqref{eq:b_tilde} does not contain the zeroth mode, this mode must be padded into the register so that the transform has the correct length. The zeroth-mode padding and the odd extension can be implemented by embedding the original register into a larger Hilbert space using ancilla qubits, followed by appropriate index-shifting operations. However, once the zeroth mode is included, the resulting transform length is no longer a power of two, but $2^{n_\mathrm{s}+1}+2$, so a standard radix-2 quantum Fourier transform is not directly applicable. We therefore use the quantum Bluestein's algorithm (QBA) introduced in Ref.~\cite{Kuo_2025}, which implements an exact discrete Fourier transform for arbitrary integer transform lengths while retaining the same asymptotic gate complexity as the radix-2 quantum Fourier transform. Similar to evaluating the measurement mismatch term in Eq.~\eqref{eq:C_BG_phys}, an additional circuit is required to measure the overlap between $\ket{\tilde{u}^\text{meas}}$ and the ansatz state after discrete sine transform, depicted in Fig.~\ref{QBA_measurement}. In total, evaluating Eq.~\eqref{eq:C_BG_sine} requires 16 circuits.

\section{Numerical simulation}
\label{sec:numerical_simulation}

\begin{table*}[t]
\caption{Summary of the numerical experiments and the RMSEs of the reconstructed data in the Galerkin-reduced-basis and the real-space approaches.}
\label{tab:summary}
\centering
\renewcommand{\arraystretch}{1.12}
\setlength{\tabcolsep}{5pt}
\begin{tabular}{l c S[table-format=1.3] c c c c c c}
\toprule
\multicolumn{1}{c}{PDE} & $u(t=0,x)$ & $\nu$ & \makecell[c]{Qubit number\\($n_{\rm t}+n_{\rm s}$)} & \makecell[c]{Number of\\sensors} 
&
$w$
&
\multicolumn{1}{c}{\makecell[c]{Measurement\\ window $t$}} &
\multicolumn{2}{c}{\makecell[c]{RMSE}} \\
\cmidrule(lr){8-9}
& & & & & & & Galerkin & Real-space \\
\midrule
Burgers &
$\sin(2\pi x)$ & 0.05 & $3+4$ & 3 &$10^{-4}$ & $0-0.175$ & $8.6 \times 10^{-4}$ & $0.0319$ \\

Burgers &
$\sin(2\pi x)$ & 0.05 & $4+4$ & 3 &$10^{-4}$ & $0-0.174$ & 0.0026 &  \\

Burgers &
$\exp[-(2\pi x-\pi)^2]$ & 0.05 & $4+4$ & 4 & $10^{-4}$ & $0-0.195$ & $0.0429$ & $0.0568$ \\    

Burgers &
$\exp[-(2\pi x-\pi)^2]$ & 0.05 & $6+6$ & 5 & $10^{-5}$ & $0-0.06$ & $0.0744$ &  \\               

Burgers &
$\exp[-(2\pi x-\pi)^2]$ & 0.008 & $3+5$ & 8 & $10^{-5}$ & $0.13-0.20$ & $0.0280$ &  \\

KS &
$\sin(2\pi x)+2\sin(\pi x)$ & 0.05 & $3+3$ & 2 &$10^{-4}$ & $0-0.056$ & $0.0089$ & $0.0267$ \\

KS &
$\sin(2\pi x)+2\sin(\pi x)$ & 0.05 & $5+3$ & 2 & $10^{-4}$ & $0-0.062$ & $0.0295$ &  \\
KS &
\makecell[c]{$\sin(2\pi x)$\\$+\,0.2\,x(1-x)(4x-1)$} & 0.008 & $3+4$ & 4 &$10^{-7}$ & $0.1465-0.15$ & $0.0454$ & \\
\bottomrule
\end{tabular}
\end{table*}

We demonstrate our method using numerical simulation for both the real-space and the Galerkin-reduced-basis approaches in the moderate-viscosity ($\nu=0.05$) and low-viscosity ($\nu=0.008$) regimes. For each approach, we validate using the simpler Burgers equation and leverage the KS equation to compare performance in a more dynamically complex system. Throughout these studies, we use only a limited number of sensors at fixed spatial locations, covering at most 25\% of the points on the discrete spacetime grid. 

For ground truth, we first simulate the velocity field $u(x,t)$ on a spacetime grid for a given initial condition $u(t=0,x)$ using fourth-order Runge–Kutta (RK4) methods. For the more complicated KS equation, the implicit–explicit RK4 (IMEXRK4)~\cite{Bhatt_2019_IMEXRK4} and the exponential time-differencing RK4 (ETDRK4) methods~\cite{Kassam_2005_ETDRK4} are used. 
This simulated result is treated as the actual values $\{u_{\mathrm{baseline}}(t,x)\}$ and provide the baseline for comparison. Then, we consider the situation where the initial condition is not accessible, and the sensors are placed at a limited set of locations $\mathcal{Q}$ at all times of the spacetime grid. The reconstruction task uses the proposed VQA based on the governing equation and these measurement data. The variational parameters are obtained using the L-BFGS-B optimizer~\cite{Byrd_1995_LBFGSB}. For each numerical experiment, we perform the optimization from 20 independent random initializations and choose the one with the lowest optimized cost function.
We first calculate the gradients for the PDE violation and the measurement mismatch in the cost function at the beginning of the optimization for all the initializations, and choose a value for the hyperparameter $w$ to balance the $\ell_2$-norms of the two gradients. As the variation in the gradient norm across different initializations is small, we use the same value for $w$ across initializations, and keep it fixed throughout each optimization.  
To represent the quality of the reconstructed solution, we use the RMSE weighted by the average magnitude of $u_{\mathrm{baseline}}$, 
\begin{equation}
    \eta = \sqrt{\frac{\mathbb{E}_{t,x} \left[ \left(u_{\mathrm{VQA}}(t,x)-u_{\mathrm{baseline}}(t,x)\right)^2 \right]}{\mathbb{E}_{t,x}\left[ u^2_{\mathrm{baseline}}(t,x) \right]}},
\end{equation}
where $u(x,t)_{\mathrm{VQA}}$ is the solution obtained from the proposed VQA and the mean is taken over all the accessible spacetime points. In the presence of the condition $u(t,x=0)=0$, we exclude the location $x=0$ from the mean since the condition is enforced by the ansatz.  
However, we remark that in reality, there usually are observables of interest other than values at all the spacetime points, such as the spatial or temporal autocorrelation of the field. The quality of the reconstructed data should be considered in the context accordingly. 
Table~\ref{tab:summary} summarizes the reconstruction settings and RMSEs for all benchmark cases.

\subsection{Moderate viscosity at $\nu=0.05$}
\label{subsec:moderate_viscosity}

\begin{figure}[t]
    \centering
    \includegraphics[width=\columnwidth]
    {\figdir/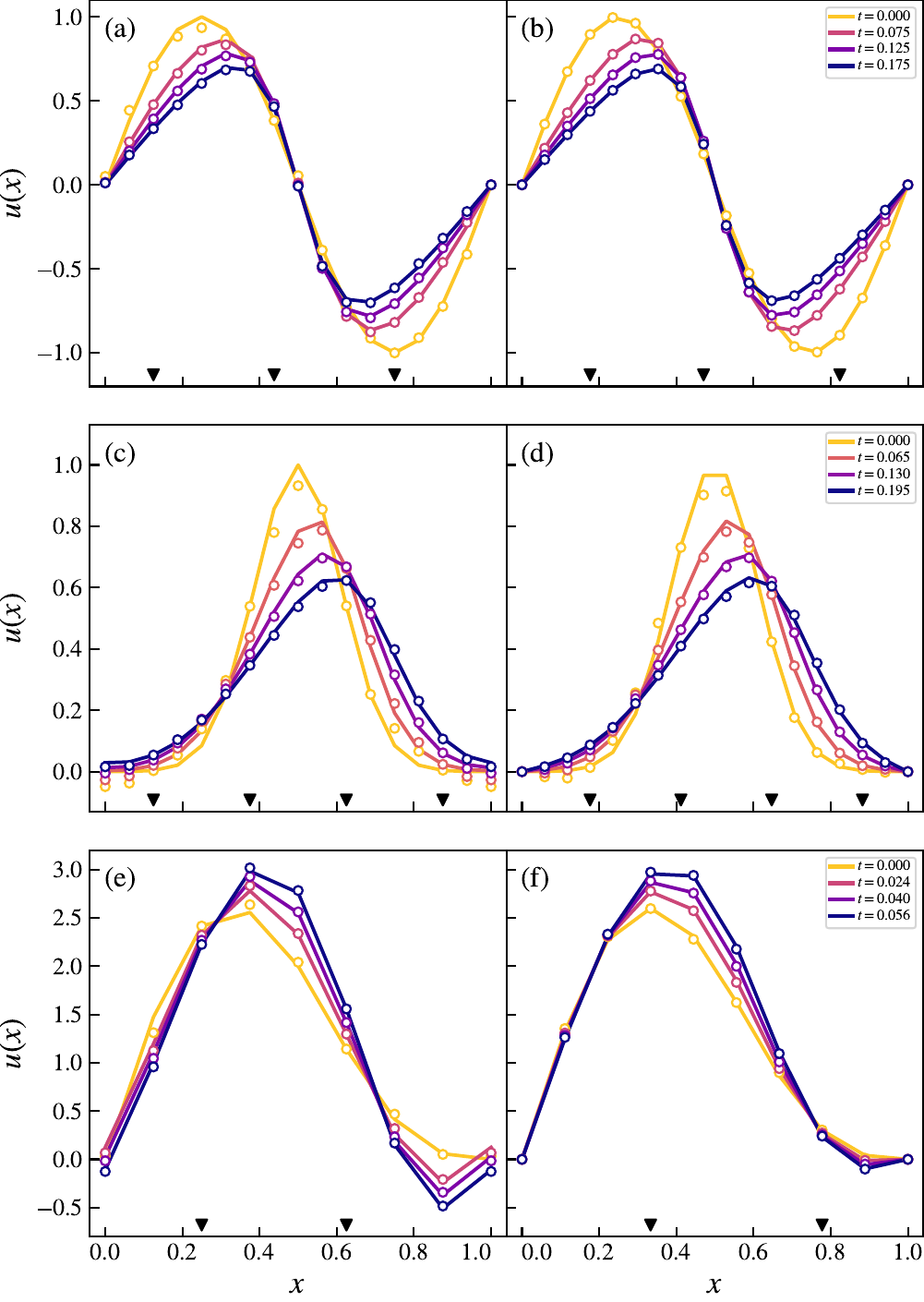}
    \caption{Numerically simulated baseline and VQA-reconstructed velocity fields with viscosity $\nu=0.05$. Triangles on the x-axis indicate the sensor locations used for the reconstruction. Each solid line connects the baseline velocities at different spatial coordinates for a given time. Unfilled circles are the velocities reconstructed by the proposed VQA. Panels (a)(b) and (c)(d) show the results for the Burgers equation with the sinusoidal and Gaussian initial wave profiles, respectively. Panels (e) and (f) show the results for the KS equation. Panels (a), (c), and (e) are results from the real-space approach and Panels (b), (d), and (f) are results from the Galerkin-reduced-basis approach.}
    \label{fig:nu0.05}
\end{figure}
In the moderate-viscosity regime, $\nu=0.05$, nonlinear advection remains visible while diffusion keeps the solution relatively smooth. Fig.~\ref{fig:nu0.05} shows the results for the Burgers equation and the KS equation using the Galerkin-reduced-basis and the real-space approaches. These correspond to the first, third, and sixth rows of data in Table~\ref{tab:summary}. For the Galerkin-reduced-basis approach with the sine basis, we enforce the additional condition $u(x=0)=u(x=L)=0$ in the baseline simulation. Setting the total length $L$ as the unit, the spatial domain is $x \in [0,1]$ with periodic boundary condition.

For the Burgers equation, the wave profile at the initial time is first set to $u(t=0,x) = \sin(2\pi x)$. We use $n_\text{t}=3$ qubits to represent time coordinates and $n_\text{s}=4$ qubits to represent either spatial coordinates (in the real-space approach) or basis modes (in the Galerkin-reduced-basis approach). Note that in the real-space representation, the spatial resolution is $\Delta x=1/16$ while in the Galerkin-reduced-basis approach, the spatial resolution is $\Delta x=1/17$ due to the truncation at $16$ modes. The temporal resolution is chosen as $\Delta t=0.025$ to maintain numerical stability (see Appendix~\ref{app:stable_condition}). The measurement window is $t=0-0.175$. 
In reconstruction, the initial wave profile is not accessible. Instead, a total of 3 sensors are placed in the spatial domain, whose locations are denoted with triangles along the $x$-axis in the plot. Each sensor records the full discrete time history of the evolution window. The real-space approach and the Galerkin-reduced-basis approach achieve an RMSE of $0.0319$ and $0.0009$, respectively. As shown in Fig.~\ref{fig:nu0.05}(a) and (b), the reconstructed profiles agree closely with the baseline solution. The variational ansatz used has 8 brickwall layers in both cases, corresponding to 104 parameters (including $\lambda_0$). 

Then we set the wave profile at the initial time to a more nontrivial form, $u(t=0,x) = \exp[-(2\pi x-\pi)^2]$, which is not sinusoidal and the condition $u(t,x=0)=0$ is only approximately met. We use $n_\text{t}+n_\text{s}=4+4$ qubits to represent the coordinates, and choose the temporal resolution $\Delta t=0.013$. The measurement window is $t=0-0.195$. With 4 sensors, the real-space approach and the Galerkin-reduced-basis approach achieve an RMSE of $0.0568$ and $0.0429$, respectively. Fig.~\ref{fig:nu0.05}(c) and (d) show the reconstructed data against the baseline solution. This demonstrates that good agreement can be reached in the Galerkin-reduced-basis approach even when the initial wave profile has a relatively complicated expansion in the chosen Galerkin basis. The variational ansatz used has 9 and 7 brickwall layers, respectively, corresponding to 135 and 107 parameters. 

For the KS equation, the initial wave profile is set to $u(t=0,x) = \sin(2\pi x) + 2\sin(\pi x)$. We use $n_\text{t}=3$ qubits to represent time coordinates and $n_\text{s}=3$ qubits to represent spatial coordinates or basis modes. The corresponding spatial resolutions are $\Delta x=1/8$ and $\Delta x=1/9$, respectively. The temporal resolution is chosen as $\Delta t=8 \times 10^{-3}$, making the measurement window $t=0-0.056$. 
2 sensors are placed at 2 interior points to record the data throughout the evolution. An ansatz with 13 brickwall layers and 137 parameters is used for the real-space approach, while an ansatz with 7 brickwall layers and 77 parameters is employed for the Galerkin-reduced-basis approach. A good overlap between the reconstructed field and the baseline is achieved, as in Fig.~\ref{fig:nu0.05}(e) and (f), with an RMSE of $0.0267$ for the real-space approach and $0.0089$ for the Galerkin-reduced-basis approach. 

For the Burgers and KS equations here, the Galerkin-reduced-basis approach achieves a better reconstruction accuracy than the real-space approach, especially when the initial wave profile is sinusoidal. Among the tested cases, the real-space approach typically requires a deeper ansatz circuit to reach a comparable accuracy level. This is likely because with this viscosity and initial wave profile, the relevant solutions admit a more compact representation in the chosen Galerkin basis, making the optimization problem better conditioned. However, we note that when the initial wave profiles are of a Gaussian form or $\sin(2\pi x)+2\sin(\pi x)$, the baseline velocity fields are slightly different for the two approaches due to the different boundary conditions at $x=0$, which do not provide a direct comparison.  

\begin{figure}[t]
    \centering
    \includegraphics[width=0.927\columnwidth]{\figdir/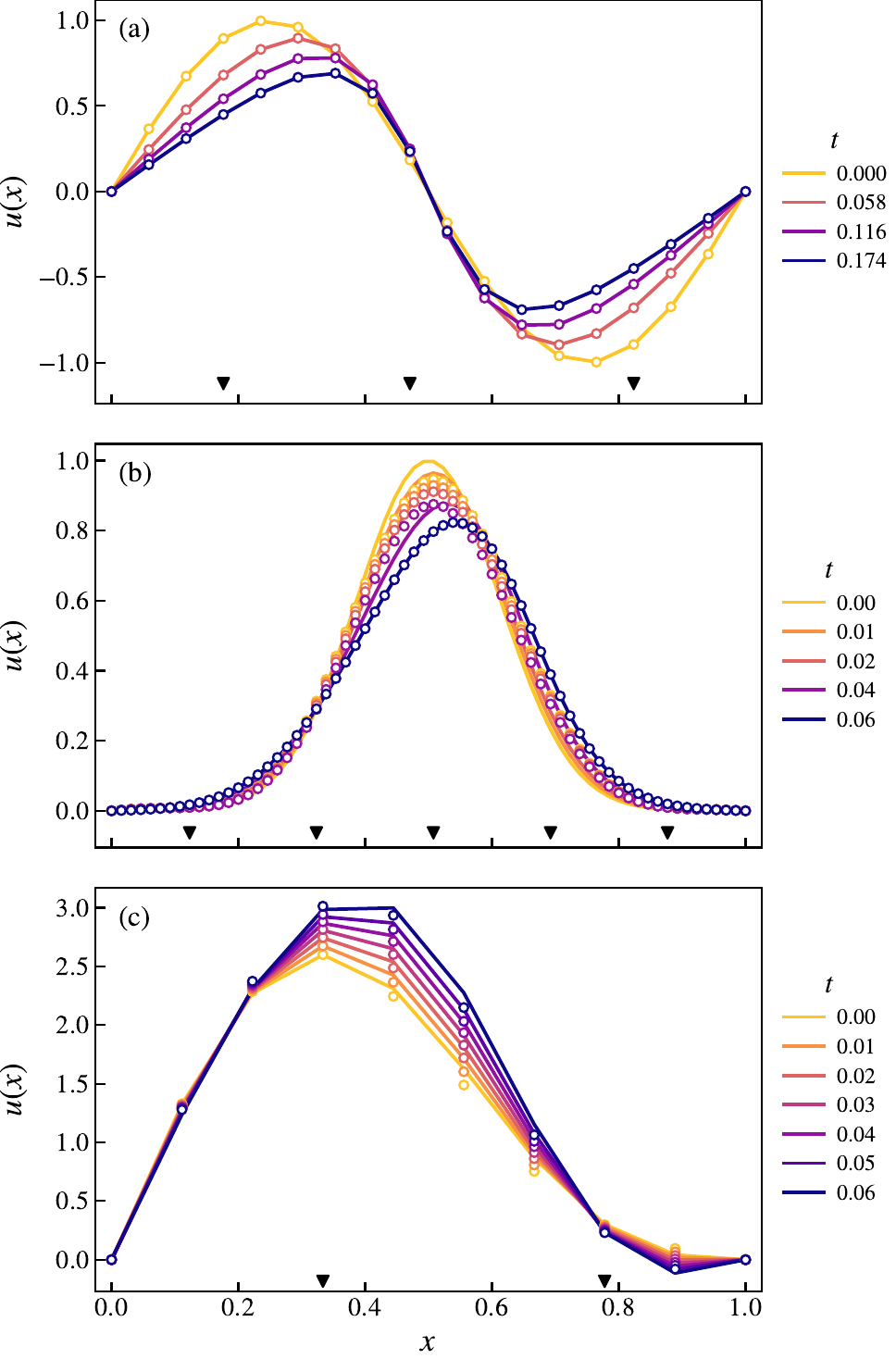}
    \caption{Numerically simulated baseline and VQA-reconstructed velocity fields with higher space or temporal resolution for $\nu=0.05$. Each solid line connects the baseline velocities at different spatial coordinates for a given time. Unfilled circles are the velocities reconstructed by the proposed VQA. Panels (a) and (b) show the results for the Burgers equation with the sinusoidal (using $n_\text{t}+n_\text{s}=4+4$ qubits) and Gaussian initial wave profiles (using $n_\text{t}+n_\text{s}=6+6$ qubits), respectively. Panel (c) shows the results for the KS equation with $n_\text{t}+n_\text{s}=5+3$ qubits.}
    \label{fig:higher_resol}
\end{figure}

\subsection{Increased resolution at $\nu=0.05$}
\label{subsec:resolution}

With the same baseline velocity fields and measurement data, we now investigate the performance of our method using the Galerkin-reduced-basis approach with an increased space or temporal resolution. The results are in Fig.~\ref{fig:higher_resol}.

For the Burgers equation with $u(t=0,x) = \sin(2\pi x)$, we use $n_\text{t}+n_\text{s}=4+4$ qubits, corresponding to an increased temporal resolution $\Delta t=0.0116$ for a similar measurement window $t=0-0.174$. 3 sensors are placed to measure the data, whose locations are marked by the triangles along the x-axis. 
Using an ansatz with $8$ brickwall layers and $121$ parameters, an RMSE of $0.0026$ is achieved. Compared to the previous $n_\text{t}+n_\text{s}=3+4$ case, the temporal resolution increases while the spatial resolution stays the same, bringing the time step further inside the stability bound. The accuracy in the best solution over 20 random initializations drops with the higher temporal resolution. The higher accuracy level in the previous case is likely a consequence of a small Hilbert space, where an 8-layer ansatz is sufficiently expressive. Now for the larger Hilbert space, we can further improve the accuracy by using a deeper ansatz circuit. With 10 brickwall layers for example, an RMSE of $0.0018$ is reached.

For the Burgers equation with the more complicated initial wave profile $u(t=0,x) = \exp[-(2\pi x-\pi)^2]$, we use $n_\text{t}+n_\text{s}=6+6$ qubits to increase the spacetime resolution to $\Delta x=1/65$ and $\Delta t=9.53\times10^{-4}$ for the measurement window $t=0-0.06$. 5 sensors provide the measurement data.
An ansatz with 4 brickwall layers and 101 parameters is sufficient to reproduce the baseline velocity field with an RMSE of $0.0744$. We compare this to the RMSE achieved in the $n_\text{t}+n_\text{s}=4+4$ case before, calculated for the restricted time window $t=0-0.065$, at the value $0.0380$. In this case the initial wave profile is less simple, and with the higher resolution in both space and time, an accuracy of the same order is achieved with a more compact ansatz circuit. We remark that the increased Hilbert space may still lead to increased difficulty in optimization.

For the KS equation, we use $n_\text{t}+n_\text{s}=5+3$ qubits, corresponding to a spatial resolution $\Delta x=1/9$ and 32 time coordinates. We take the measurement window $t=0-0.062$ with a step $\Delta t=2\times10^{-3}$. 2 sensors measure the data used for reconstruction. In this case, an ansatz with 7 brickwall layers and 107 parameters reconstruct a velocity field in good agreement with the baseline values, achieving an RMSE of $0.0295$. 
To compare it to the RMSE achieved in the $n_\text{t}+n_\text{s}=3+3$ case before, we restrict the time window to $t=0-0.056$, giving an RMSE of $0.0274$. Now with the increased temporal resolution only, the accuracy is lowered, likely due to the increased challenges in optimization.

Additional numerical simulation results for different initial wave profiles and spacetime resolutions are presented in Appendix~\ref{app:additional_results}.

\subsection{Low viscosity at $\nu=0.008$}
\label{subsec:low_viscosity}

\begin{figure}[t]
    \centering
    \includegraphics[width=0.8\columnwidth]
{\figdir/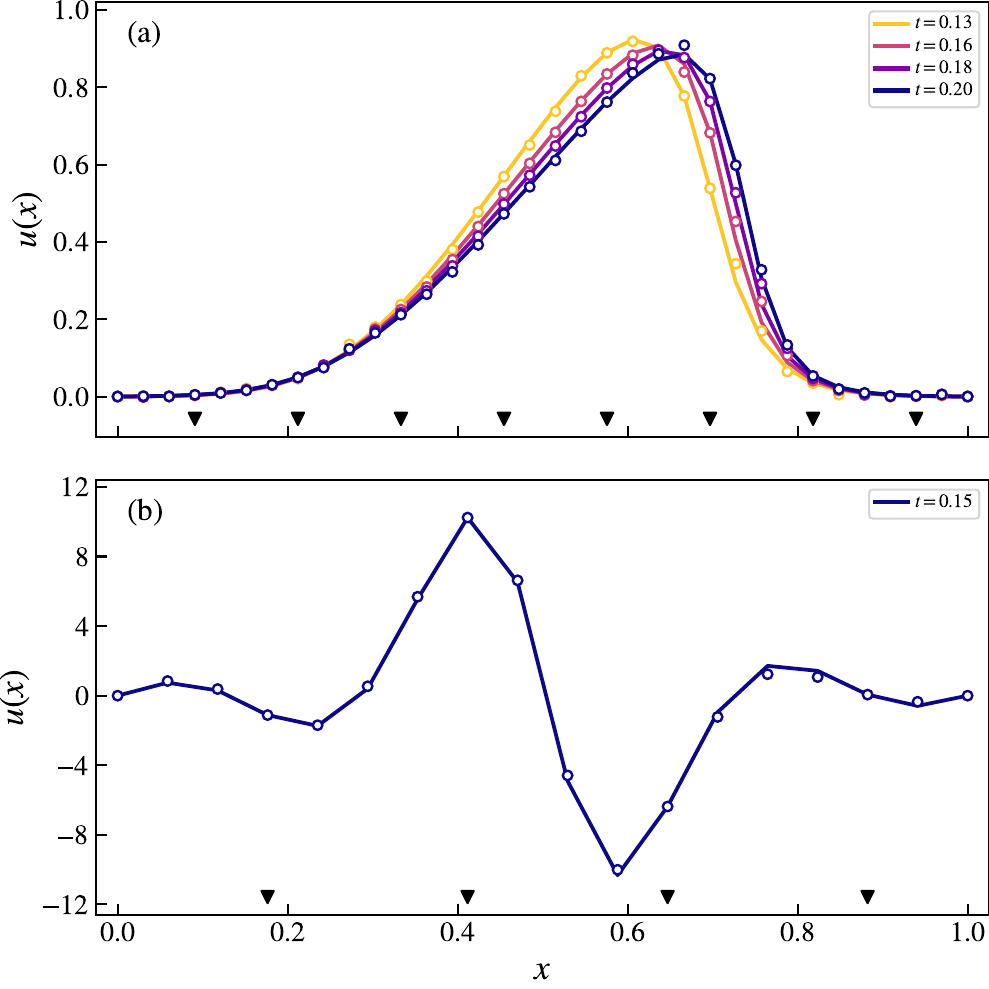}
    \caption{
    Numerically simulated baseline velocities with viscosity $\nu=0.008$ and reconstructed velocities using the Galerkin-reduced-basis approach. Triangles on the x-axis indicate the sensor locations used for the reconstruction. Each solid line connects the baseline velocities at different spatial coordinates for a given time. Unfilled circles are the velocities reconstructed by the proposed VQA. Panel (a) shows the results for the Burgers equation. Panel (b) shows the wave profile at the end of the evolution window for the KS equation.
    }
    \label{fig:nu0.008}
\end{figure}
 
In many experiments, measurements may begin at arbitrary times without a well-defined initial condition. We therefore consider a more challenging setting: reconstructing velocity fields over arbitrary time windows in the low-viscosity regime $\nu=0.008$. We model this situation by simulating the baseline velocities from $t=0$ and selecting a measurement window at later times. In this regime, the wave develops steeper gradients during the evolution and the optimization is oblivious to the wave profile at the initial time.
We focus on the Galerkin-reduced-basis approach in this challenging setting as it provides a suitable basis for representing the solution. Fig.~\ref{fig:nu0.008} shows the results for the Burgers and the KS equations, corresponding to the fifth and eighth rows of data in Table~\ref{tab:summary}.

For the Burgers equation, we focus on the case with a nontrivial initial wave profile of a Gaussian form $u(t=0,x)=\exp[-(2 \pi x-\pi)^2]$. The selected measurement window is $t=0.13-0.20$, with a step size $\Delta t=0.01$. We use $n_\text{t}=3$ and $n_\text{s}=5$ qubits. A finer spatial grid is required to resolve the steeper gradients of the wave profile compared to the moderate viscosity case. 
Using an ansatz with 9 brickwall layers, the reconstructed wave profiles remain in close agreement with the baseline values throughout the selected measurement time window, as shown in Fig.~\ref{fig:nu0.008}(a). Despite the sharper waveform structure formed in the front of the wave during this time, the VQA reconstruction preserves both the location and the shape of the evolving peak, yielding an RMSE of $0.0280$.

For the KS equation, we take an initial wave profile that deviates from a sinusoidal form, $u(t=0,x)=\sin(2\pi x)+0.2x(1-x)(4x-1)$, to test the utility of the method. 
We use $n_\text{t}=3$ and $n_\text{s}=4$ qubits, and an ansatz with 10 brickwall layers. The temporal grid is much finer than that for the Burgers equation because the KS equation leads to more complex dynamics. With a step size $\Delta t=5\times10^{-4}$, the measurement window is $t=0.1465-0.15$. We choose this small measurement window due to the limited computational resources in our simulation. A larger number of time points, determined by $n_\text{t}$, leads to a larger Hilbert space that needs to be simulated. Since the measurement window is chosen to be small, we only show the wave profile at the final time $t=0.15$ in Fig.~\ref{fig:nu0.008}(b). Even in this challenging case, the reconstructed velocities stay aligned with the baseline values, yielding an RMSE of $0.0454$.

These results demonstrate that with a suitable representation of the solution (the Galerkin-reduced-basis approach here), we can reconstruct the velocity field in the low-viscosity regime with the presence of steep gradients, with a delayed measurement window. This is more relevant to real experiments where data may be collected at arbitrary times. 

\section{Conclusions}
\label{sec:conclusion}

In this work, we develop a VQA for reconstructing velocity fields governed by nonlinear fluid equations from sparse sensors. The solution on a spacetime grid is encoded in the amplitudes of a variational quantum state that can be prepared on a quantum device with tunable circuit parameters. The qubits are partitioned into two sets, denoting the time coordinate and the spatial distribution, respectively. We take two approaches corresponding to different ways of representing the spatial distribution of the solution field. In the real-space approach, the computational basis states represent a discrete set of spatial coordinates. This spacetime encoding has proved useful in finding the solution on a spacetime grid with a known initial condition~\cite{Pool_2024}.
In the Galerkin-reduced-basis approach, the computational basis states represent a finite set of Galerkin basis chosen for the problem. In our examples, the Burgers and the KS equations in a one-dimensional periodic space with a special condition $u(x=0)=0$, the Galerkin basis is chosen to be a truncated series of sine terms. The solution is obtained by varying the parameters to optimize a cost function, which contains one term corresponding to violation of the governing PDE equation and one term corresponding to mismatch with the measured data from the sensors.

Given the variational state prepared on a quantum device, the cost function can be measured using quantum circuits based on QNPU. The number of circuits is constant in the number of qubits while the extraction circuit (excluding the ansatz) depth scales linearly. For the variational state, we take a hardware-efficient ansatz containing single-qubit rotations and two-qubit entangling gates with a brickwall layout. This kind of ansatz has been adopted in optimization and ground-state finding problems, including variationally solving PDE with a known initial condition, due to its compatibility with near-term quantum devices~\cite{sarma2024quantum, Pool_2024, Pia2026}. Such problem-agnostic ansatze are known to face challenges in optimization when the system size is large, and incorporating problem-specific information into the ansatz will likely improve the performance of the algorithm~\cite{Grimsley2019, Tang2021, Grimsley2023, Stadelmann2025adapt}. We leave the problem of constructing a problem-informed variational ansatz for future work.

We demonstrate our method for the Burgers and the KS equations in a one-dimensional periodic space, providing the circuits to measure the cost functions and numerically simulating the performance for two viscosity settings. In our numerical experiment, the ideal values for the velocity field are simulated using RK4 methods, from which the measurement data are extracted for a limited set of sensor locations (up to 25\% of the spatial coordinates). The VQA-reconstructed velocity fields show good agreement with the baseline values.

Our method provides a framework for utilizing quantum devices to reconstruct nonlinear PDE fields from sparse measurements in experiments. We show that encoding the solution on a spacetime grid naturally fits the reconstruction setup and extend the spatial encoding beyond the values on discrete spatial coordinates to a Galerkin basis suitable for the problem. We construct the cost function based on both the PDE equation and the measurement data, and show how to design the circuits for measuring the cost function. This significantly expands the capabilities of traditional diagnostic methods, which typically provide only sparse point measurements or partial information (e.g., a single velocity component), by enabling the reconstruction of the complete field.

At the same time, several important questions remain open. In the presence of noise on quantum devices, there are errors associated with the measured cost function. It is unclear how such errors, as well as the errors in the experimental measurement data from the sensors, affect the reconstruction accuracy. Moreover, the optimization efficiency is crucial in the success of the VQA approach, and is affected by the sparsity of the measurement data. Efficient optimization also demands a good ansatz as mentioned above. 
Finally, velocities at an exponentially large set of spacetime coordinates cannot be efficiently extracted~\cite{HHL, Aaronson2015, Montanaro2016}. Direct extraction of physically relevant observables from the reconstructed fields, rather than full tomography of the reconstructed solution state, should be considered when investigating possible quantum advantage over classical approaches~\cite{Montanaro2016, Linden2022, Bravyi2025, Bravyi2026}. These considerations can motivate several directions for future work, including larger-scale benchmarks with higher dimensions, noisy reconstruction under realistic sparse measurement data with an efficient readout scheme to extract useful physical observables.


\begin{acknowledgements}
The authors gratefully acknowledge the financial support from AFOSR under the grant number FA9550-25-1-0029. In addition, N.-Q. N. and Y. C. acknowledge support from startup funds provided by Florida State University. 
\end{acknowledgements}


\appendix

\section{Cost functions for the KS equation}
\label{app:KS_cost}

The cost functions for the KS equation in the real-space and the Galerkin-reduced-basis approaches are, respectively,
\begin{align}
\label{C_KS_phys}
    \mathcal{C}_{\mathrm{KS}}^{\mathrm{real}} &= \lambda_0^2 w \sum_{i=0}^{2^{n_\mathrm{t}}-2} \sum_{j=0}^{2^{n_\mathrm{s}}-1}
    [\tilde{u}_{i+1,j} - \tilde{u}_{i,j} \nonumber\\
    &+\Delta t (\lambda_0\tilde{u}_{i,j}\nabla \tilde{u}_{i,j}+\Delta \tilde{u}_{i,j}+\nu \Delta^2 \tilde{u}_{i,j})]^2 \nonumber\\
    &+\sum_{i=0}^{2^{n_\mathrm{t}}-1}\ \sum_{j\in \mathcal{Q}}
    \left[\lambda_0 \tilde{u}_{i,j} - \lambda^{\mathrm{meas}} \tilde{u}^{\mathrm{meas}}_{i,j}\right]^2,
\end{align}
and
\begin{align}
\label{C_KS_sine}
    \mathcal{C}_{\mathrm{KS}}^{\mathrm{Galerkin}}
    &= \lambda_0^2 w \sum_{i=0}^{2^{n_\mathrm{t}}-2} \sum_{j=0}^{2^{n_\mathrm{s}}-1}\Bigg\{\tilde{b}_{i+1,j} - \tilde{b}_{i,j} \nonumber\\
    &+ \Delta t \Bigg[ \left(-\left(\frac{(j+1)\pi}{L}\right)^2 + \nu \left(\frac{(j+1)\pi}{L}\right)^4\right) \tilde{b}_{i,j} \nonumber\\
    &-\frac{\lambda_0}{\sqrt{2(N+1)}} \frac{(j+1)\pi}{L} \sum_{k=0}^{N-j-2}\tilde{b}_{i,k}\tilde{b}_{i,k+j+1} \nonumber\\ 
    &+\frac{\lambda_0}{2\sqrt{2(N+1)}} \frac{(j+1)\pi}{L}\sum_{k=0}^{j-1} \tilde{b}_{i,k}\tilde{b}_{i,j-k-1}\Bigg] \Bigg\}^2 \nonumber\\
    &+\sum_{i=0}^{2^{n_\mathrm{t}}-1}\ \sum_{j\in \mathcal{Q}} \left[u({i \Delta{t},j \Delta{x}}) - \lambda^{\mathrm{meas}} \tilde{u}^{\mathrm{meas}}_{i,j}\right]^2 .
\end{align}
Here the velocity field is related to the state amplitudes according to Eq.~\eqref{eq:u_tilde} in the real-space approach, and Eqs.~\eqref{eq:b_tilde} and \eqref{eq:DST} in the Galerkin-reduced-basis approach, for a system with a total of $n_\text{t}+n_\text{s}$ qubits. A single cost-function evaluation requires $19$ distinct measurement circuits in the real-space approach and $21$ in the Galerkin-reduced-basis approach.

\section{Choice of time step}
\label{app:stable_condition}
In our approach, Eq.~\eqref{eq:Cpde} shows the PDE violation in the cost function is constructed using a forward-Euler discretization in time. For the advection term $u\partial u_x$ in Eqs.~\eqref{eq:BG_eqn} and \eqref{eq:KS_eqn}, the Courant-Friedrichs-Lewy (CFL) condition provides a temporal-resolution estimate \cite{LeVeque_2007_CFL}
\begin{equation}
\label{CFL}
\begin{split}
&\Delta t_{\mathrm{adv}} \lesssim \frac{\Delta x}{\text{max} |u(x,t)|},
\end{split}
\end{equation}
In the real-space approach, the spatial derivatives are approximated using second-order central difference. For the Burgers equation, the diffusion term $\nu\partial_x^2u$ leads to a diffusive stability limit from the von Neumann stability analysis~\cite{LeVeque_2007_CFL}. For the KS equation, the term $\nu\partial_x^4u$ gives rise to a hyperdiffusive stability limit with the same mechanism~\cite{Strikwerda_2004_Diffusion}. The corresponding estimates can be written as     
\begin{equation}
\label{Diffusive_stability_limit}
\begin{split}
\Delta t_{\mathrm{diff}} &\lesssim \frac{\Delta x^2}{2 \nu},\\
\Delta t_{\mathrm{hyperdiff}} &\lesssim \frac{\Delta x^4}{8 \nu}.
\end{split}
\end{equation}
In the Galerkin-reduced-basis approach where a sine basis is used, the corresponding explicit time-step estimates are instead written in terms of the largest resolved wavenumber $k_{\max}=\mathcal{O}(1/\Delta x)$, leading to the same scaling $\Delta t \lesssim (\nu k_{\max}^2)^{-1}$ for diffusive stability in the Burgers equation and $\Delta t \lesssim (\nu k_{\max}^4)^{-1}$ for hyperdiffusive stability in the KS equation. These are conservative worst-case estimates for explicit time marching.

Eq.~(\ref{CFL}) and (\ref{Diffusive_stability_limit}) are treated as estimates based on worst-case explicit-Euler stability. In our simulations, we use them to guide the choice of the time step. To observe evolution of the waveform, we stay close to the bound. Near this worst-case bound, if we add one more qubit to resolve the spatial coordinates, it requires approximately 2 more qubits for the Burgers equation and 4 more qubits for the KS equation to maintain a compatible temporal resolution.

\section{Additional simulation results}
\label{app:additional_results}

\begin{table*}[t]
\caption{Summary of the additional numerical experiments and the RMSEs of the reconstructed data in the Galerkin-reduced-basis and the real-space approaches.}
\label{tab:summary_appendix}
\centering
\renewcommand{\arraystretch}{1.12}
\setlength{\tabcolsep}{5pt}
\begin{tabular}{l c S[table-format=1.3] c c c c c c}
\toprule
\multicolumn{1}{c}{PDE} & $u(t=0,x)$ & $\nu$ & \makecell[c]{Qubit number\\($n_t+n_s$)} & \makecell[c]{Number of \\ sensors} 
&
$w$
&
\multicolumn{1}{c}{\makecell[c]{Measurement \\ window $t$}} &
\multicolumn{2}{c}{\makecell[c]{RMSE}} \\
\cmidrule(lr){8-9}
& & & & & & & Galerkin & Real-space \\
\midrule
Burgers &
$\exp[-(2\pi x-\pi)^2]$ & 0.05 & $5+4$ & 4 & $10^{-4}$ & $0-0.4$ & $0.0436$ &  \\
                   
Burgers &
$\exp[-(2\pi x-\pi)^2]$ & 0.05 & $5+5$ & 5 & $10^{-6}$ & $0-0.14$ & $0.0300$ &  \\

Burgers &
$\sin(2\pi x)$ & 0.008 & $3+5$ & 8 & $10^{-5}$ & $0.278-0.285$ & $0.0161$ & \\

KS &
$\sin(2\pi x)$ & 0.05 & $3+3$ & 2 & $10^{-4}$ & $0-0.07$ & $0.0107$ & $0.0329$  \\

KS & 
$\sin(2\pi x)$ & 0.008 & $3+4$ & 4 & $10^{-7}$ & $0.1465-0.1500$ & $0.0062$ & \\
\bottomrule
\end{tabular}
\end{table*}

\begin{figure}[t]
    \centering
\includegraphics[width=0.927\columnwidth]{\figdir/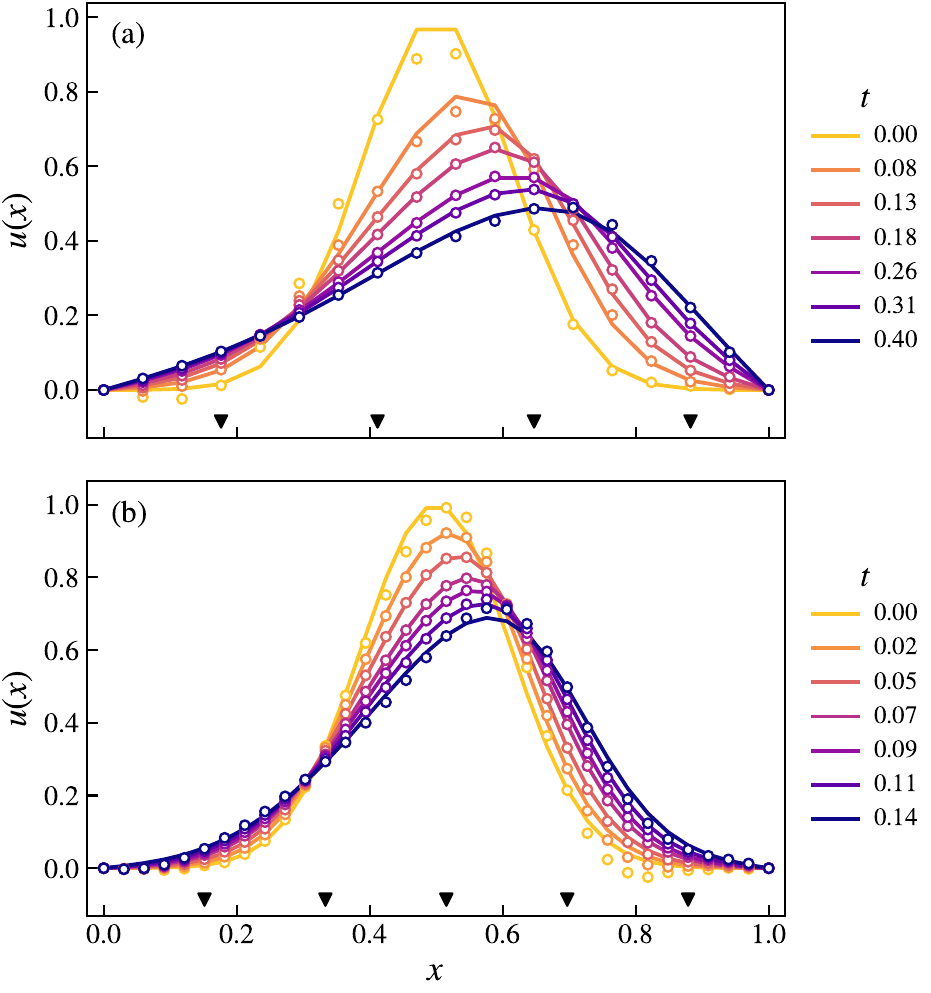}
    \caption{Numerically simulated baseline and VQA-reconstructed velocity fields with different spacetime resolutions for the Burgers equation with $\nu = 0.05$. Each solid line connects the baseline velocities at different spatial coordinates for a given time. Unfilled circles are the velocities reconstructed by the proposed VQA. Panels (a) and (b) show the results with $n_\text{t} + n_\text{s} =5+4$ qubits and $n_\text{t} + n_\text{s} = 5+5$ qubits, respectively.
    }
    \label{fig:BG_gaussian}
\end{figure}

Table~\ref{tab:summary_appendix} summarizes additional numerical simulation results. In all our simulations, we set the domain length $L$ as the unit. 

For the Burgers equation, Fig.~\ref{fig:BG_gaussian} shows the reconstructed velocity fields for the same initial wave profile ($u(t=0,x)=\exp[-(2 \pi x -\pi)^2]$) and same viscosity ($\nu =0.05$) as in the main text with different spacetime resolutions. First we extend the spacetime grid to $n_\text{t} + n_\text{s} = 5+4$. Setting the same $\Delta t= 0.013$, this allows for a larger measurement window $t=0-0.4$. With 4 sensors, an ansatz with 7 brickwall layers and $122$ parameters reaches an RMSE of $0.0436$. The accuracy for this larger measurement window is comparable to the level achieved in the $n_\text{t} + n_\text{s} = 4+4$ case. Then we set $n_\text{t} + n_\text{s} = 5+5$, and a higher temporal resolution $\Delta t=4.52 \times 10^{-3}$ which gives the measurement window $t=0-0.14$. With 5 sensors, an ansatz with 7 brickwall layers and $137$ parameters achieves an RMSE of $0.0300$, slightly better than the level in the $n_\text{t} + n_\text{s} = 4+4$ case.

\begin{figure}[h]
    \centering
    \includegraphics[width=\columnwidth]{\figdir/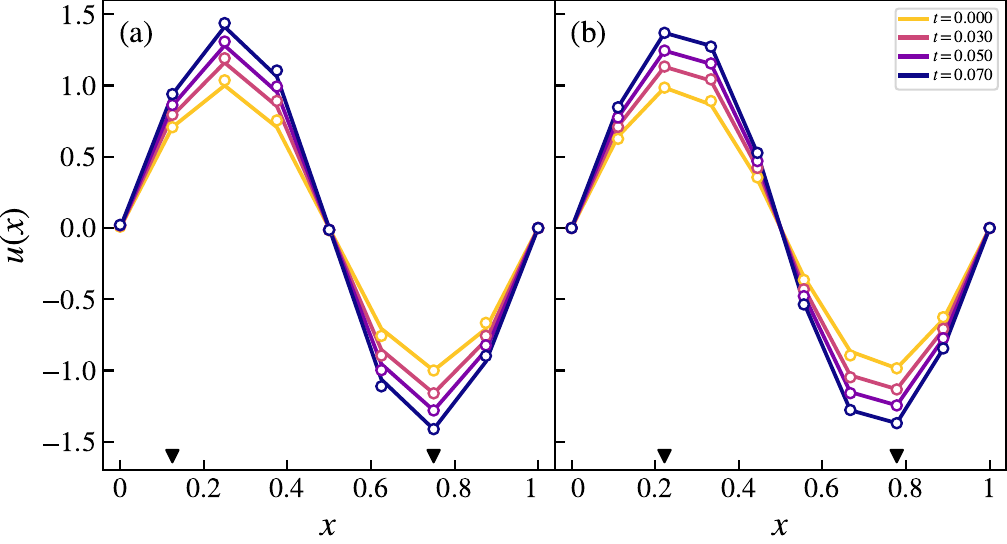}
    \caption{Numerically simulated baseline and VQA-reconstructed velocity fields for the KS equation with viscosity $\nu =0.05$. Each solid line connects the baseline velocities at different spatial coordinates for a given time. Unfilled circles are the velocities reconstructed by the proposed VQA. Panels (a) and (b) show the results in the real-space and the Galerkin-reduced-basis approaches, respectively.
    }
    \label{fig:KS_sine}
\end{figure}

\begin{figure}[h]
    \centering
\includegraphics[width=0.8\columnwidth]{\figdir/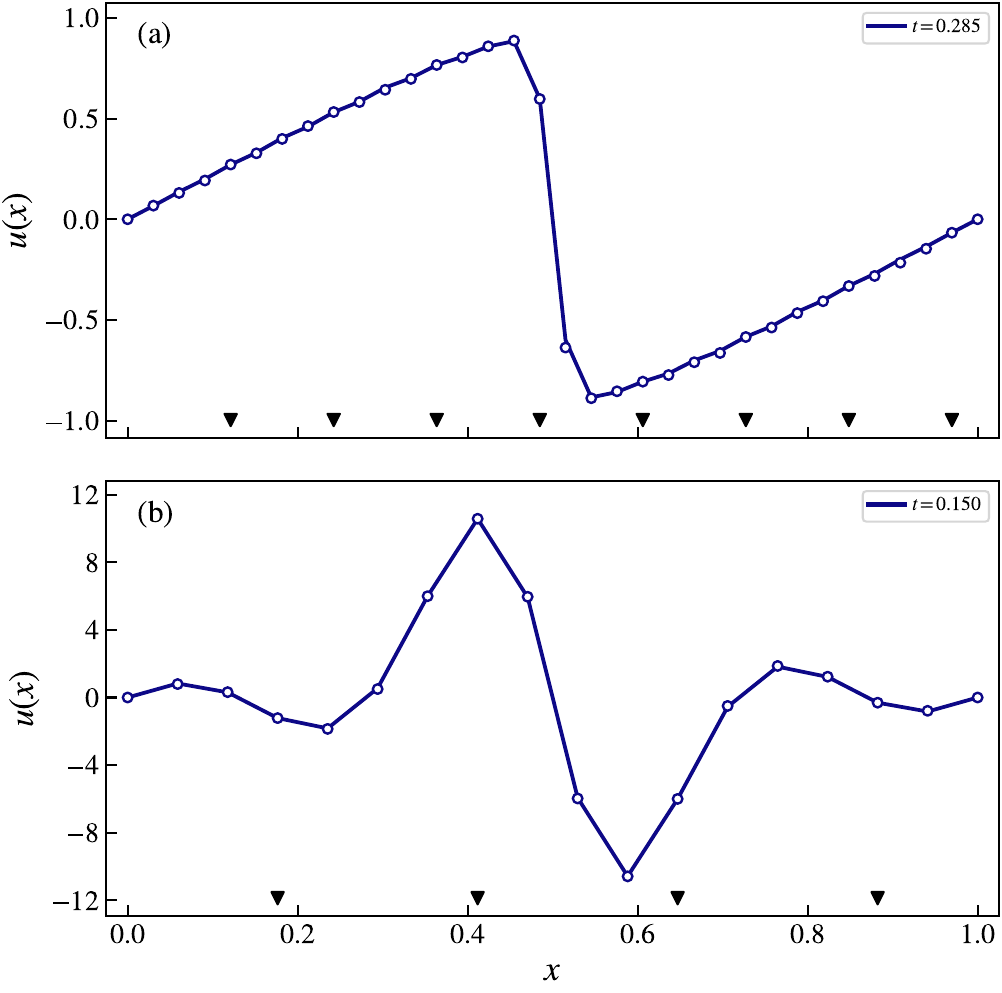}
    \caption{Numerically simulated baseline velocities at the end of a measurement window and the reconstructed velocities using the Galerkin-reduced-basis approach, in the case $\nu = 0.008$ and $u(t=0,x)=\sin(2\pi x)$. Each solid line connects the baseline velocities at different spatial coordinates for a given time. Unfilled circles are the velocities reconstructed by the proposed VQA. Panels (a) and (b) show the final wave profiles for the Burgers and the KS equation, respectively.}
    \label{fig:extra_nu0.008}
\end{figure}

For the KS equation, we simulate the reconstruction with a simpler initial wave profile $u(t=0,x)=\sin(2\pi x)$ than that in the main text. An $n_\text{t}+n_\text{s} = 3+3$-qubit encoding is employed for the measurement window $t=0-0.07$, with $\Delta t=0.01$ and 2 sensors collecting measurement data. Using an ansatz with 13 brickwall layers and 137 parameters, we obtain an RMSE of $0.0329$ in the real-space approach and an RMSE of $0.0107$ in the Galerkin-reduced-basis approach. Fig.~\ref{fig:KS_sine} shows that the reconstructed data from both approaches agree closely with the baseline values and yield RMSEs of the same magnitude order as in the main text where the initial wave profile is more complicated.

Fig.~\ref{fig:extra_nu0.008} shows additional final-time wave profiles in the small viscosity regime $\nu =0.008$ with a measurement window excluding $t=0$ for Burgers and KS equations, respectively. The baseline velocity simulation starts from a simple initial wave profile $u(t=0,x)=\sin(2\pi x)$. For the Burgers equation, an $n_\text{t}+n_\text{s}=3+5$-qubit encoding is employed for the measurement window $t=0.278-0.285$, with a timestep $\Delta t=10^{-3}$. Despite the presence of a steep gradient around $x=0.5$, an ansatz with 7 brickwall layers is able to reconstruct the velocity field with a low RMSE of $0.0161$. For the KS equation, we use an $n_\text{t}+n_\text{s}=3+4$-qubit encoding for the measurement window $t=0.1465-0.15$, with $\Delta t = 5 \times 10^{-4}$. Similarly, a steep gradient is formed around $x=0.5$, and an ansatz with 8 brickwall layers obtains a low RMSE of $0.0062$. 


\bibliographystyle{apsrev4-1}
\bibliography{references}

\end{document}